\documentclass[10pt]{article}
\usepackage{titlesec}
\usepackage{amsmath,esdiff}
\usepackage{amssymb}
\usepackage{authblk}
\usepackage[normalem]{ulem}
\usepackage{bbold}
\usepackage{geometry}
\usepackage{setspace}
\usepackage{cite}
\usepackage[colorlinks=true,linkcolor=blue,citecolor=red,urlcolor=red]{hyperref}%
\usepackage{mathrsfs}
\usepackage{hyperref}
\usepackage{graphicx}
\usepackage{cancel}
\usepackage{array,esint}
\usepackage[most]{tcolorbox}
\titleformat{\section}{\fontsize{12}{12}\bfseries}{\thesection}{1em}{}
\graphicspath{}
\begin{document}
\title{\textbf{Path integral action for a resonant detector of gravitational waves in the generalized uncertainty principle framework}}
\author{\textbf{$\mathbf{Soham}$ $\mathbf{Sen}^{a}$\thanks{sensohomhary@gmail.com}, $\mathbf{Sukanta}$ $\mathbf{Bhattacharyya}^{b}$\thanks{sukanta706@gmail.com} and $\mathbf{Sunandan}$ $\mathbf{Gangopadhyay}^{a}$\thanks{sunandan.gangopadhyay@gmail.com}}}
\affil{{${}^a$ Department of Astrophysics and High Energy Physics,}\\
{S.N. Bose National Centre for Basic Sciences}\\
{JD Block, Sector III, Salt Lake, Kolkata 700106, India}\\
{${}^b$ Department of Physics,}\\ {West Bengal State University, Barasat, Kolkata 700126, India}}
\date{}
\maketitle
\begin{abstract}
\noindent The Heisenberg uncertainty principle gets modified by the introduction of an observer independent minimal length. In this work we have considered the resonant gravitational wave detector in the modified uncertainty principle framework where we have used the position momentum uncertainty relation with a quadratic order correction only. We have then used the path integral approach to calculate an action for the bar detector in presence of a gravitational wave and then derived the Lagrangian of the system leading to the equation of motion for the configuration-space position coordinate in one dimension. We then find a perturbative solution for the coordinate of the detector for a circularly polarized gravitational wave leading to a classical solution of the same for given initial conditions. Using this classical form of the coordinate of the detector, we finally obtain the classical form of the on-shell action describing the harmonic oscillator-gravitational wave system. Finally, we have obtained the free particle propagator containing the quantum fluctuation term considering gravitational wave interaction.
\end{abstract}
\section{Introduction}
Quantum mechanics and general relativity are the two most successful theories explaining the phenomena at two most fundamental length scales of the universe. While quantum mechanics explains the intricacies of the atomic length scale, general relativity \cite{Einstein15,Einstein16} sheds light on the large-scale structure of the universe. In order to understand the fundamental mysteries of the universe, we need a quantum theory of gravity, explaining the analytical structure of the gravitational interaction at the quantum length scale. Theories like loop quantum gravity\cite{ROVELLI,CARLIP}, string theory\cite{ACV,KOPAPR}, and noncommutative geometry\cite{GLO} have provided a convincing theoretical framework explaining the Planck-scale nature of gravity but none of them have compelling experimental evidence to support their claim of providing an exact description of the quantum nature of gravity.  Although all of them prescribe the existence of an observer-independent minimal length which can be incorporated by the modification of the standard Heisenberg uncertainty principle (HUP), also known as the generalized uncertainty principle (GUP). The first few attempts to improvise an integral relation between minimal length scale and gravity was shown in \cite{BRON1,BRON2}, followed by \cite{MEAD}. We also get a strong evidence of the existence of this fundamental length scale from the various gedanken experiments in quantum gravity phenomenology as well.  This GUP framework has been used to investigate several areas of theoretical physics, like black hole physics and its thermodynamics\cite{MAGGIORE,SCARDIGLI,ADLERSAN,ADLERCHENSAN,RABIN,SG1,SCARDIGLI2,
SG2,Ong,EPJC,BMajumder}, various quantum systems like particle in a box, simple harmonic oscillator\cite{DAS1,DAS2}, optomechanical systems\cite{IVASP,IVASP2,KSP}, and gravitational wave bar detectors\cite{SG3,SG4}. There have been several recent studies involving the path integral formalism of a non-relativistic particle moving in an arbitrary potential in the generalized uncertainty principle framework\cite{SG3,DAS3,SGSB}. The simplest form the modified HUP can be written in the following form\cite{Kempf}
\begin{equation}\label{1.01}
\Delta q_i\Delta p_i\geq \frac{\hbar}{2}\left[1+\gamma\left(\Delta p^2+\left<p\right>^2\right)+2\gamma\left(\Delta p_i^2+\left<p_i\right>^2\right)\right];~i=1,2,3
\end{equation}
where $p^2=\sum_{k=1}^3p_kp_k$ and $q_k, p_k$ are the phase space position and its conjugate momenta.
In eq.(\ref{1.01}) the GUP parameter $\gamma$ in terms of the dimensionless parameter $\gamma_0$ can be recast as follows
\begin{equation}\label{1.02}
\gamma=\frac{\gamma_0}{m_p^2 c^2}
\end{equation}
where $m_p$ is the Planck mass and $c$ is the speed of light. It is quite natural to realize that the order of magnitude of the GUP parameter will play a significant role in providing an understanding of the GUP effects. There have been several studies to find a bound on the GUP parameter itself\cite{SG4,SCARDIGLI2, DAS1, BM, FENG, BBGK, SCARDIGLI3,GRD,RCSG,OTM}. 

\noindent In the 1969, the first proposition to detect gravitational waves was made by J. Weber\cite{Detect1} which was followed by a subsequent paper in 1982 by Ferrari \textit{et} \textit{al.}\cite{Detect2}. Bar detectors currently have a sensitivity $\frac{\Delta L}{L}\sim 10^{-19}$\cite{Detect3}, where $\Delta L$ is fractional variation of the length $L$ ($\sim1$ meter) of the bar detector. A historical perspective of these resonant detectors is given in \cite{Aguiar}. The detection of gravitational waves by the LIGO\cite{LIGO,LIGO2} and Virgo\cite{VIRGO} detectors have unveiled a new realm of quantum gravity phenomenology. There have been several recent investigations regarding the traces of quantum gravitational effects in these gravitational wave detectors. A lot of investigation has been done to check if any signature of this fundamental Planck length, whether it is noncommutativity\cite{SGA1,SGA2,SGA3,SGA4,SGA5,SGA6} or GUP\cite{LIGO3,SG4}, is visible in GW bar detectors. We would like to point out that till date there has not been a successful detection of gravitational waves in resonant bar detectors. However, there is a good hope that the sensitivity of the detectors would increase in future enabling detection of these waves. The AURIGA (\textit{Antenna Ultracriogenica Risonante per l'Indagine Gravitazionale Astronomica}) detector at INFN, Italy is probably the only functional bar detector. These bar detectors are sensitive to frequencies of the order of $1$kHz along with a strain sensitivity of the order $10^{-19}$\cite{Piz}. In case of astrophysical events, collapsing and bouncing cores of supernova can be a source to huge intensities of gravitational waves having frequencies in the vicinity of $1-3$kHz. The value of the strain sensitivity can be calculated using Thorne's formula\cite{Thorne}. The strain sensitivity ($h$), according to this formula, is given by
\begin{equation}\label{0.01}
h=2.7\times 10^{-20}\left[\frac{\Delta E_{GW}}{M_sc^2}\right]^{\frac{1}{2}}\left[\frac{1kHz}{f}\right]^{\frac{1}{2}}\left[\frac{10Mpc}{d}\right]
\end{equation}
where $\Delta E_{GW}$ is the energy converted to gravitational waves, $f$ is the characteristic frequency of the burst, $M_s$ is the solar mass and $d$ is the distance of the burst source from earth. A possible value of fraction of energy converted to gravitational waves for supernova events is around $7\times10^{-4}$. Now for $h\sim 3\times 10^{-19}$ and $f\sim 0.9 kHz$, the distance $d$ has the value around $25kpc$. The occurrence of such a supernova event of the required magnitude at this distance from the earth would definitely result in the detection of gravitational waves by the bar detectors. An effort to increase the sensitivity of these detectors to $h\sim10^{-20}$ is presently being carried out, and achieving this sensitivity would increase the distance of the supernova event from the earth to $250kpc$, which is more likely to occur. The main motivation to work with a gravitational wave bar detector is that it is a very resourceful and economic alternative to the LIGO/VIRGO detectors. 


\noindent In this work we investigate the path integral formalism of a resonant gravitational wave bar detector interacting with the gravitational wave emitted from a distant source in the GUP framework. The incoming gravitational waves interact with the elastic matter in the resonant bar detector causing tiny vibrations called phonons. Physically we can describe these detectors as a quantum mechanical \textit{gravitational wave-harmonic oscillator} (GW-HO) system because we call these vibrations; the quantum mechanical forced harmonic oscillator. To calculate the perturbative solution to the system we use the gravitational wave and generalized uncertainty modifications as perturbations. Our study presents a path integral approach to look at such a system and is the first work using path integral. The advantage of working with path integrals is that the effective action describing the system can be easily read off from the structure of the configuration space path integral\cite{SGPRL}. 
\section{The gravitational wave resonant detector interaction model}
\noindent To begin the discussion, we need to write down the Hamiltonian for the resonant bar detector in the presence of a gravitational wave in the generalized uncertainty principle framework. The modified commutation relation following from eq.(\ref{1.01}) takes the following from\cite{Kempf} 
\begin{equation}\label{1.03}
[\hat{q}_i,\hat{p}_j]=i\hbar\left[\delta_{ij}+\gamma\delta_{ij}\hat{p}^2+2\gamma\hat{p}_i\hat{p}_j\right]
\end{equation}
where $i,j=1,2,3$. The modified position and momentum operators $\hat{q}_i$ and $\hat{p}_i$ in terms of the usual variables $\hat{q}_{0i}$ and $\hat{p}_{0i}$ read
\begin{equation}\label{1.04}
\hat{q}_i=\hat{q}_{0i}~,~\hat{p}_i=\hat{p}_{0i}\left(1+\gamma \hat{p}_0^2\right)~.
\end{equation}
Here $\hat{p}_0^2=\sum_{k=1}^3\hat{p}_{0k}\hat{p}_{0k}$ and $[\hat{q}_{0i},\hat{p}_{0j}]=i\hbar\delta_{ij}$.
In order to write the Hamiltonian of the system we start by analyzing the background metric as a superposition of a small perturbation on the flat background metric. The background metric is taken as follows
\begin{equation}\label{1.05}
g_{\mu\nu}=\eta_{\mu\nu}+h_{\mu\nu}
\end{equation}
where $\eta_{\mu\nu}=\text{diag}\{1,-1,-1,-1\}$ and $|h_{\mu\nu}|\ll 1$.   We now consider a two dimensional harmonic oscillator with mass $m$ and intrinsic frequency $\varpi$. The geodesic deviation equation for the aforementioned system in the proper detector frame is given as follows\cite{Maggiore}
\begin{equation}\label{1.06}
\begin{split}
m\ddot{q}^k&=-mR^k_{~0l0}q^l-m\varpi^2q^k\\
\implies \ddot{q}^k&=\frac{d\Gamma^k_{~0l}}{dt}q^l-\varpi^2q^k~;~k=1,2
\end{split}
\end{equation}
where $R^k_{~0l0}$ in terms of the background perturbation is given by
\begin{equation}\label{1.07}
R^k_{~0l0}=-\frac{d\Gamma^k_{~0l}}{dt}=-\frac{\ddot{h}_{kl}}{2}~.
\end{equation}
Note that here we are using the transverse traceless gauge to get rid of the unphysical degrees of freedom. The Lagrangian from which eq.(\ref{1.07}) can be obtained reads
\begin{equation}\label{1.08}
L=\frac{1}{2}m\dot{q}_k^2-m\Gamma^{k}_{~0l}\dot{q}_kq^l-\frac{1}{2}m\varpi^2q_k^2~.
\end{equation}
The Hamiltonian corresponding to the Lagrangian in eq.(\ref{1.08}) reads
\begin{equation}\label{1.09}
H=\frac{1}{2m}\left(p_k+m\Gamma^k_{~0l}q^l\right)^2+\frac{1}{2}m\varpi^2q_l^2~.
\end{equation}
To write the Hamiltonian in eq.(\ref{1.09}) in quantum mechanical description we just elevate $q$ and $p$ to the operator prescription. Therefore, the Hamiltonian in terms of the position and momentum operators can be expressed as follows

\begin{equation}\label{1.10}
\hat{H}=\frac{1}{2m}\left(\hat{p}_k+m\Gamma^k_{~0l}\hat{q}^l\right)^2+\frac{1}{2}m\varpi^2\hat{q}_l^2~.
\end{equation} 
Using the representation of the position and momentum operators in eq.(\ref{1.04}), the Hamiltonian (\ref{1.10}) {of the GW-HO system in presence of GUP } can be written as follows
\begin{equation}\label{1.11}
\hat{H}=\left(\frac{\hat{p}_{0k}^2}{2m}+\frac{1}{2}m\varpi^2\hat{q}_{0k}^2\right)+\frac{\gamma}{m}\hat{p}_{0k}^2\hat{p}_0^2+\frac{1}{2}\Gamma^{k}_{~0l}\left(\hat{p}_{0k}\hat{q}^{0l}+\hat{q}^{0l}\hat{p}_{0k}\right)+\frac{\gamma}{2}\Gamma^{k}_{~0l}\left(\hat{p}_{0k}\hat{p}_0^2\hat{q}^{0l}+\hat{q}^{0l}\hat{p}_{0k}\hat{p}_0^2\right)~.
\end{equation}
Now a typical bar is a cylinder of length $L \equiv 3 m$ and radius $R \equiv 30 cm$\cite{Maggiore}. Hence in a first approximation, we can treat the GW detector in presence of GUP as a one dimensional HO. The Hamiltonian in eq.(\ref{1.11}) can be recast in one dimension as follows
\begin{equation}\label{1.12}
\hat{H}=\frac{p^2}{2m}+\frac{1}{2}m\varpi^2q^2+\gamma\frac{p^4}{m}+\frac{1}{2}\Gamma^1_{~01}(pq+qp)+\frac{\gamma}{2}\Gamma^1_{~01}(p^3q+qp^3)
\end{equation} 
where for notational simplicity we have used $\hat{p}_{01}=p$ and $\hat{q}_{01}=q$. In the next section we will proceed to construct the path integral formalism of the GW-HO system in presence of the GUP and calculate the propagation kernel for that system.
\section{Path integral and the propagation kernel}
In this section we will use the Hamiltonian in eq.(\ref{1.12}) to calculate the propagation kernel via the path integral approach. We consider the initial and the final state of the Hamiltonian in eq.(\ref{1.12}) at initial time $t_i$ and final time $t_f$ as $\left|q_i,t_i\right>$ and $\left|q_f,t_f\right>$ respectively. The general form of the propagation kernel can be written as follows
\begin{equation}\label{1.13}
\begin{split}
\left<q_f,t_f|q_i,t_i\right>&=\lim\limits_{N\rightarrow\infty}\int_{-\infty}^{+\infty}dq_{N-1}\ldots dq_1\left<q_f,t_f|q_{N-1},t_{N-1}\right>\left<q_{N-1},t_{N-1}|q_{N-2},t_{N-2}\right>\ldots\left<q_1,t_1|q_{i},t_{i}\right>\\
&=\lim\limits_{N\rightarrow\infty}\int_{-\infty}^{+\infty}\prod\limits_{\alpha=1}^{N-1}dq_\alpha\left<q_f\right|e^{-\frac{i\hat{H}(t_f-t_{N-1})}{\hbar}}\left|q_{N-1}\right>\ldots\left<q_1\right|e^{-\frac{i\hat{H}(t_1-t_i)}{\hbar}}\left|q_i\right>\\
&=\lim\limits_{N\rightarrow\infty}\int_{-\infty}^{+\infty}\prod\limits_{\alpha=1}^{N-1}dq_\alpha\prod\limits_{\beta=0}^{N-1}\left<q_{\beta+1}\right|e^{-\frac{i\hat{H}(t_{\beta+1}-t_\beta)}{\hbar}}\left|q_\beta\right>
\end{split}
\end{equation}
where $t_f=t_N$, $t_i=t_0$ and $t_{N}-t_{N-1}=\Delta t$. Now we will introduce the complete set of momentum eigenstates $\left(\int_{-\infty}^{+\infty}dp\left|p\right>\left<p\right|=1\right)$ in the following way
\begin{equation}\label{1.14}
\begin{split}
\left<q_f,t_f|q_i,t_i\right>=&\lim\limits_{N\rightarrow\infty}\int\prod\limits_{\alpha=1}^{N-1}dq_\alpha\prod\limits_{\beta=0}^{N-1}\int ~dp_\beta\left<q_{\beta+1}|p_\beta\right>\left<p_\beta|q_\beta\right>\exp\left(-\frac{iH(q_\beta,p_\beta)(t_{\beta+1}-t_\beta)}{\hbar}\right)\\
=&\lim\limits_{N\rightarrow\infty}\int_{-\infty}^{+\infty}\prod\limits_{\alpha=1}^{N-1}dq_\alpha\prod\limits_{\beta=0}^{N-1}\int_{-\infty}^{+\infty} \frac{dp_\beta}{2\pi\hbar}\exp\biggr[\frac{i\Delta t}{\hbar}\sum\limits_{\beta=0}^{N-1}\biggr[\frac{p_\beta(q_{\beta+1}-q_\beta)}{\Delta t}-\biggr(\frac{p_\beta^2}{2m}+\frac{1}{2}m\varpi^2q_\beta^2+\frac{\gamma p_\beta^4}{m}\\&+\frac{p_\beta q_\beta(h_{\beta+1}-h_\beta)}{2\Delta t}+\frac{\gamma p_\beta^3q_\beta({h}_{\beta+1}-h_\beta)}{2\Delta t}\biggr)\biggr]\biggr]
\end{split}
\end{equation}
where we have used $h_{11}=h$~. The final form of eq.(\ref{1.14}) in the $\Delta t\rightarrow0$ limit can be recast as follows
\begin{equation}\label{1.15}
\left<q_f,t_f|q_i,t_i\right>=\int\mathcal{D}q\mathcal{D}p\exp\left(\frac{i}{\hbar}\mathcal{S}\right)
\end{equation}
where $\mathcal{S}$ is the phase space action. The phase space action is given as follows
\begin{equation}\label{1.16}
\mathcal{S}=\int_{t_i}^{t_f} dt\left[p\dot{q}-\left(\frac{p^2}{2m}+\frac{\dot{h}_{11}}{2}pq+\frac{1}{2}m\varpi^2q^2+\frac{\gamma p^4}{m}+\frac{\gamma\dot{h}_{11}}{2}p^3q\right)\right]~.
\end{equation}
To obtain the configuration space Lagrangian we will simplify eq.(\ref{1.14}) as follows
\begin{equation}\label{1.17}
\begin{split}
\left<q_f,t_f|q_i,t_i\right>&\cong\lim\limits_{N\rightarrow\infty}\int_{-\infty}^{+\infty}\prod\limits_{\alpha=1}^{N-1}dq_\alpha\prod\limits_{\beta=0}^{N-1}\int_{-\infty}^{+\infty} \frac{dp_\beta}{2\pi\hbar}\left[1-\frac{i\gamma\Delta t}{m\hbar}\left(p_\beta^4+\frac{h_{\beta+1}-h_\beta}{2\Delta t}p_\beta^3q_\beta\right)+\mathcal{O}(\gamma^2)\right]
\end{split}
\end{equation}
\begin{equation*}
\begin{split}
\times\exp\left[\frac{i\Delta t m}{2\hbar}\left[\left(\frac{q_{\beta+1}-q_\beta}{\Delta t}-\frac{h_{\beta+1}-h_\beta}{4\Delta t}q_\beta\right)^2-\varpi^2q_\beta^2\right]\right]\exp\left[-\frac{i\Delta t}{2m\hbar}\left[p_\beta-\left(\frac{m(q_{\beta+1}-q_\beta)}{\Delta t}-\frac{m(h_{\beta+1}-h_\beta)q_\beta}{4\Delta t}\right)\right]^2\right]~.
\end{split}
\end{equation*}
To do the momentum integral for each $\beta$ value, we shall do the following coordinate transformation
\begin{equation}\label{1.18}
\bar{p}_\beta=p_\beta-\left(\frac{m(q_{\beta+1}-q_\beta)}{\Delta t}-\frac{m(h_{\beta+1}-h_\beta)q_\beta}{4\Delta t}\right)~.
\end{equation}
Using eq.(\ref{1.18}) in eq.(\ref{1.17}), the propagation kernel upto $\sim \gamma,h$  can be recast as
\begin{equation}\label{1.19}
\begin{split}
\left<q_f,t_f|q_i,t_i\right>\cong&\lim\limits_{N\rightarrow\infty}\int_{-\infty}^{+\infty}\prod\limits_{\alpha=1}^{N-1}dq_\alpha\prod\limits_{\beta=0}^{N-1}\int_{-\infty}^{+\infty} \frac{d\bar{p}_\beta}{2\pi\hbar}\left[1-\frac{i\gamma\Delta t}{m\hbar}\biggr[\biggr(\bar{p}_\beta+\left(\frac{m(q_{\beta+1}-q_\beta)}{\Delta t}-\frac{m(h_{\beta+1}-h_\beta)q_\beta}{4\Delta t}\right)
\right)^4+\\&\frac{h_{\beta+1}-h_\beta}{2\Delta t}\left(\bar{p}_\beta+\left(\frac{m(q_{\beta+1}-q_\beta)}{\Delta t}-\frac{m(h_{\beta+1}-h_\beta)q_\beta}{4\Delta t}\right)\right)^3q_\beta\biggr]+\mathcal{O}(\gamma^2)\biggr]\exp\left[-\frac{i\Delta t}{2m\hbar}{\bar{p}_\beta}^2\right]\\&\times\exp\left[\frac{i\Delta t m}{2\hbar}\left[\left(\frac{q_{\beta+1}-q_\beta}{\Delta t}-\frac{h_{\beta+1}-h_\beta}{4}q_\beta\right)^2-\varpi^2q_\beta^2\right]\right]~.
\end{split}
\end{equation}
The momentum integral in eq.(\ref{1.19}) can be obtained as follows
\begin{equation}\label{1.20}
\begin{split}
\left<q_{\beta+1},t_{\beta+1}|q_\beta,t_\beta\right>\cong&\sqrt{\frac{m}{2\pi i\hbar\Delta t}}\biggr\{1-6\gamma m^2\left(\frac{q_{\beta+1}-q_\beta}{\Delta t}-\frac{h_{\beta+1}-h_\beta}{4\Delta t}q_\beta\right)^2-\frac{3\gamma m^2(h_{\beta+1}-h-\beta)}{2\Delta t}\biggr(\frac{q_{\beta+1}-q_\beta}{\Delta t}\\&-\frac{(h_{\beta+1}-h_\beta)}{4\Delta t}q_\beta\biggr)q_\beta+\frac{3i\gamma m\hbar}{\Delta t}\biggr\}~\exp\biggr[\frac{im\Delta t}{2\hbar}\biggr[\biggr(\frac{q_{\beta+1}-q_\beta}{\Delta t}-\frac{(h_{\beta+1}-h_\beta)q_\beta}{4}\biggr)^2-2\gamma m^2\\&\times\biggr[\biggr(\frac{q_{\beta+1}-q_\beta}{\Delta t}-\frac{(h_{\beta+1}-h_\beta)q_\beta}{4}\biggr)^4+\frac{(h_{\beta+1}-h_\beta)q_\beta}{2}\biggr(\frac{q_{\beta+1}-q_\beta}{\Delta t}-\frac{(h_{\beta+1}-h_\beta)q_\beta}{4}\biggr)^3\biggr]-\varpi^2q_\beta^2\biggr]\biggr]~.
\end{split}
\end{equation}
Using eq.(\ref{1.20}) in eq.(\ref{1.19}), we obtain the form of the propagation kernel upto some constant factor as follows
\begin{equation}\label{1.21}
\begin{split}
\left<q_f,t_f|q_i,t_i\right>&=\int_{-\infty}^{+\infty}\prod\limits_{\alpha=1}^ {N-1}dq_\alpha\exp\biggr[\sum\limits_{\beta=0}^{N-1}\frac{im\Delta t}{2\hbar}\biggr\{\biggr(\frac{q_{\beta+1}-q_\beta}{\Delta t}-\frac{(h_{\beta+1}-h_\beta)q_\beta}{4}\biggr)^2-2\gamma m^2\biggr(\biggr(\frac{q_{\beta+1}-q_\beta}{\Delta t}\\&-\frac{(h_{\beta+1}-h_\beta)q_\beta}{4}\biggr)^4+\frac{(h_{\beta+1}-h_\beta)q_\beta}{2}\biggr(\frac{q_{\beta+1}-q_\beta}{\Delta t}-\frac{(h_{\beta+1}-h_\beta)q_\beta}{4}\biggr)^3\biggr)-\varpi^2q_\beta^2\biggr\}\biggr]~.
\end{split}
\end{equation}
Imposing the $\Delta t\rightarrow 0$ limit in eq.(\ref{1.21}), the final form of the propagation kernel has the usual configuration space path integral structure as follows
\begin{equation}\label{1.22}
\left<q_f,t_f|q_i,t_i\right>=\mathcal{N}(T,\gamma,\dot{h})\int\mathcal{D}qe^{\frac{i}{\hbar}S}~.
\end{equation}
In the above equation the configuration space structure of the action $S$ is given as follows
\begin{equation}\label{1.23}
\begin{split}
S=&\int_{t_i}^{t_f}dt\left(\frac{m}{2}\left(\dot{q}-\frac{\dot{h}q}{4}\right)^2-\frac{1}{2}m\varpi^2q^2-\gamma m^3\left(\dot{q}-\frac{\dot{h}q}{4}\right)^4-\frac{\gamma m^3\dot{h}q}{2}\left(\dot{q}-\frac{\dot{h}q}{4}\right)^3\right)\\
\cong&\int_{t_i}^{t_f}dt\left(\frac{m}{2}\dot{q}^2-\frac{1}{2}m\varpi^2q^2-\frac{m\dot{h}\dot{q}q}{4}-\gamma m^3\dot{q}^4+\frac{1}{2}m^3\gamma\dot{h}\dot{q}^3q\right)~.
\end{split}
\end{equation}
In the last line of the above eq.(\ref{1.23}), we have kept terms upto $\mathcal{O}(h,\gamma)$.
The Lagrangian can be easily read off from eq.(\ref{1.23}) as follows
\begin{equation}\label{1.24}
L=\frac{m}{2}\dot{q}^2-\frac{1}{2}m\varpi^2q^2-\frac{m\dot{h}\dot{q}q}{4}-\gamma m^3\dot{q}^4+\frac{1}{2}\gamma m^3\dot{h}\dot{q}^3q~.
\end{equation}
The equation of motion following from the Lagrangian reads
\begin{equation}\label{1.25}
\ddot{q}-\frac{\ddot{h}q}{4}+\varpi^2 q-12m^2\gamma\ddot{q}\dot{q}^2+3\gamma m^2\dot{h}\ddot{q}\dot{q}q+\frac{3}{2}\gamma m^2\ddot{h}\dot{q}^2q+\gamma m^2\dot{h}\dot{q}^3=0~.
\end{equation}
In the next section we calculate the classical solution for the above equation of motion.
\section{Obtaining the classical solution for a periodic circularly polarized gravitational wave}
To obtain the classical solution we shall consider a circularly polarized gravitational wave in the transverse traceless gauge.  Now for a periodic circularly polarized gravitational wave the perturbation term $h$ containing the polarization information reads
\begin{equation}\label{1.26}
h_{kl}(t)=2f_0\left(\varepsilon_\times(t)\sigma^1_{kl}+\varepsilon_+(t)\sigma^3_{kl}\right)~;~k,l=1,2
\end{equation}
where $2f_0$ is the amplitude of the gravitational wave (here $f_0$ is very small), $\sigma^1$ and $\sigma^3$ are the Pauli spin matrices. In eq.(\ref{1.26}) $(\varepsilon_+(t),\varepsilon_\times(t))$ are the two possible polarization states of the gravitational wave satisfying the condition $\varepsilon_+(t)^2+\varepsilon_\times(t)^2=1$. In this particular scenario the chosen functional forms of the polarization states can be written as follows
\begin{equation}\label{1.27}
\varepsilon_+(t)=\cos(\Omega t)~,~\varepsilon_\times(t)=\sin(\Omega t)
\end{equation}
with $\Omega$ being the frequency of the gravitational wave. In our case, we will consider that the only non zero polarization state is $\varepsilon_+(t)=\cos(\Omega t)$. Therefore, in one dimension the perturbation term can be written as $h=2f_0\cos(\Omega t)$. The equation of motion in eq.(\ref{1.25}) upto $\mathcal{O}(f_0,\gamma)$ takes the form as follows
\begin{equation}\label{1.27a}
\ddot{q}+\omega^2q-12m^2\gamma\ddot{q}\dot{q}^2=0
\end{equation}
where $\omega^2=\varpi^2-\frac{\ddot{h}}{4}$. For the equation of motion in eq.(\ref{1.27a}), we consider a solution upto $\mathcal{O}(f_0,\gamma)$ as
\begin{equation}\label{1.28}
q(t)=q_0(t)+f_0 q_{f_0}(t)+\gamma q_\gamma(t)~.
\end{equation}
For the form $q(t)$ in the above equation, we obtain the solution of eq.(\ref{1.25}) as a linear combination as $q_0(t), q_{f_0}(t) \text{ and }q_\gamma(t)$. The analytical forms of $q_0(t), q_{f_0}(t) \text{ and }q_\gamma(t)$ are given as follows 
\begin{align}
q_0(t)=&\mathcal{A}_1\cos(\varpi t)+\mathcal{A}_2\sin(\varpi t)~,\label{1.29}\\
q_{f_0}(t)=&\mathcal{A}_3\cos(\varpi t)+\mathcal{A}_4\sin(\varpi t)-\frac{\Omega}{2(4\varpi^2-\Omega^2)}[\Omega\cos(\Omega t)\left\{\mathcal{A}_1\cos(\varpi t)+\mathcal{A}_2\sin(\varpi t)\right\}\nonumber\\&-2\varpi\sin(\Omega t)\left\{\mathcal{A}_2\cos(\varpi t)-\mathcal{A}_1\sin(\varpi t)\right\}]~,\label{1.30}\\
q_\gamma(t)=&\mathcal{A}_5\cos(\varpi t)+\mathcal{A}_6\sin(\varpi t)-\frac{3m^2\varpi^2}{2}[t\varpi \mathcal{A}_1(\mathcal{A}_1^2+\mathcal{A}_2^2)\sin(\varpi t)-t\varpi \mathcal{A}_2(\mathcal{A}_1^2+\mathcal{A}_2^2)\cos(\varpi t)\nonumber\\&+\frac{\mathcal{A}_1}{4}(\mathcal{A}_1^2-3\mathcal{A}_2^2)\cos(3\varpi t)-\frac{\mathcal{A}_2}{4}(\mathcal{A}_2^2-3\mathcal{A}_1^2)\sin(3\varpi t)]\label{1.31}
\end{align}
where $\mathcal{A}_1,\mathcal{A}_2,\mathcal{A}_3,\mathcal{A}_4,\mathcal{A}_5\text{ and }\mathcal{A}_6$ are arbitrary constants which we will calculate for the $q_{cl}(t)$. To obtain the form of the above constants we will apply the following set of the initial conditions
\begin{equation}\label{1.32}
\begin{split}
q(t)=\biggr\{\begin{matrix}
&q_0&\text{for}&t=0\\
&q_f&\text{for}&t=T
\end{matrix}~.
\end{split}
\end{equation}
Using the initial conditions in eq.(\ref{1.32}), the constants can be obtained as follows
\begin{align}
\mathcal{A}_1&=q_0~,~\mathcal{A}_2=\frac{q_f-q_0\cos(\varpi T)}{\sin(\varpi T)}~,\label{1.33}\\
\mathcal{A}_3&=\frac{\mathcal{A}_1\Omega^2}{2(4\varpi^2-\Omega^2)}~,\label{1.34}\\
\mathcal{A}_4&=\frac{\Omega\left\{\cos(\varpi T)\left[\Omega\mathcal{A}_1\cos(\Omega T)-2\varpi\mathcal{A}_2\sin(\Omega T)\right]+\sin(\varpi T)\left[\Omega\mathcal{A}_2\cos(\Omega T)+2\varpi \mathcal{A}_1\sin(\omega T)\right]\right\}}{2(4\varpi^2-\Omega^2)\sin(\varpi T)}-\mathcal{A}_3\cot(\varpi T)~,\label{1.35}\\
\mathcal{A}_5&=\frac{3}{8}m^2\varpi^2\mathcal{A}_1\left(\mathcal{A}_1^2-3\mathcal{A}_2^2\right)~,\label{1.36}\\
\mathcal{A}_6&=\frac{3m^2\varpi^2\left[\varpi T\left(\mathcal{A}_1\sin(\varpi T)-\mathcal{A}_2\cos(\varpi T)\right)(\mathcal{A}_1^2+\mathcal{A}_2^2)+ \frac{\mathcal{A}_1(\mathcal{A}_1^2-3\mathcal{A}_2^2)\cos(3\varpi T)}{4}-\frac{\mathcal{A}_2(\mathcal{A}_2^2-3\mathcal{A}_1^2)\sin(3\varpi T)}{4}\right]}{2\sin(\varpi T)}-\mathcal{A}_5\cot(\varpi T).\label{1.37}
\end{align} 
Using eq.(\ref{1.28}) along with eq.(s)(\ref{1.33}-\ref{1.37}) in eq.(\ref{1.23}) (with $h$ being replaced by $2f_0\cos(\Omega t)$), we obtain the form of the classical action upto $\mathcal{O}(\gamma,f)$ as follows

\begin{equation}\label{1.39}
S_\mathcal{C}=S_{\mathcal{C}}^{(0)}+S_{\mathcal{C}}^{(\gamma)}+S_{\mathcal{C}}^{(f_0)}
\end{equation}
where $S_{\mathcal{C}}^{(0)},~S_{\mathcal{C}}^{(\gamma)}$, and $S_{\mathcal{C}}^{(f_0)}$ are given by the following equations
\begin{align}
S_{\mathcal{C}}^{(0)}&=\frac{m\varpi}{2\sin(\varpi T)}\left((q_0^2+q_f^2)\cos(\varpi T)-2q_0q_f\right)~,\label{1.40}\\
S_{\mathcal{C}}^{(\gamma)}&=-\frac{\gamma m^3\varpi^3}{32\sin^4(\varpi T)}\biggr[12\varpi T \left(q_f^4+4q_f^2q_0^2+q_0^4\right)-48q_0q_f\varpi T\cos(\varpi T)(q_f^2+q_0^2)+24 q_0^2q_f^2\varpi T \cos(2\varpi T)\nonumber\\&-44q_0q_f\sin(\varpi T)(q_0^2+q_f^2)+4\sin(2\varpi T)\left(2q_0^4+15q_0^2q_f^2+2q_f^4\right)-12q_0q_f\sin(3\varpi T)(q_0^2+q_f^2)+\sin(4\varpi T)(q_0^4+q_f^4)\biggr],\label{1.41}
\end{align}
\begin{align}\label{1.41a}
S_{\mathcal{C}}^{(f_0)}&=-\frac{f_0m\varpi\Omega}{2\sin(\varpi T)(4\varpi^2-\Omega^2)}\biggr[\frac{\varpi\sin(\Omega T)}{\sin(\varpi T)}\left(q_0^2-2q_0q_f\cos(\varpi T)+q_f^2\cos(2\varpi T)\right)+2q_0q_f\Omega\cos^2\left(\frac{\Omega T}{2}\right)\nonumber\\
&-\Omega\cos(\varpi T)\left(q_0^2+q_f^2\cos(\Omega T)\right)\biggr]~. 
\end{align}
Therefore, we now have the final form of the propagator for the resonant bar detector interacting with a gravitational wave as follows
\begin{equation}\label{1.42}
\left<q_f,T|q_0,0\right>=\sqrt{\frac{m\varpi}{2\pi i\hbar \sin(\varpi T)}}\tilde{\mathcal{N}}(T,\gamma,f_0)e^{\frac{i}{\hbar}S_{cl}}.
\end{equation}
To obtain an overall structure of the fluctuation parameter in the above equation we consider the free particle structure involving gravitational wave (GW) interaction only. In this case, the infinitesimal propagator considering the particle GW interaction from eq.(\ref{1.14}) can be extracted as follows (in the $\varpi\rightarrow 0$ limit)
\begin{equation}\label{1.43}
\begin{split}
\left<q_1,\Delta t|q_0,0\right>=&\int_{-\infty}^{\infty}\frac{dp_0}{2\pi\hbar}\exp\biggr[\frac{i\Delta t}{\hbar}\left(p_0\frac{(q_1-q_0)}{\Delta t}-\left(\frac{p_0^2}{2m}+\frac{\gamma p_0^4}{m}+\frac{p_0q_0f_0}{\Delta t}\left(\cos(\Omega\Delta t)-1\right)\right)\right)\biggr]\\
\simeq&\sqrt{\frac{m}{2\pi i\hbar\Delta t}}e^{\frac{im}{2\hbar \Delta t}(q_1-	q_0)^2}\biggr[1+\frac{3im\gamma\hbar}{\Delta t}-6\gamma m^2\left(\frac{q_1-q_0}{\Delta t}\right)^2-\frac{i\gamma m^3(q_1-q_0)^4}{\hbar\Delta t^3}\\&-\frac{if_0q_0}{\hbar}\left(\frac{m(q_1-q_0)}{\Delta t}\right)(\cos(\Omega\Delta t)-1)\biggr]~.
\end{split}
\end{equation}
Now the total propagator can be written using the set of infinitesimal propagators as follows
\begin{equation}\label{1.44}
\begin{split}
\left<q_f,T|q_0,0\right>&\simeq\left(\frac{m}{2\pi i\hbar\Delta t}\right)^\frac{N}{2}\int dq_1dq_2\cdots dq_{N-1} e^{\frac{im}{2\hbar\Delta t}\left[(q_1-q_0)^2+(q_2-q_1)^2+\cdots+(q_f-q_{N-1})^2\right]}\biggr[1+\frac{3i\gamma m\hbar N}{\Delta t}-\frac{6\gamma m^2}{\Delta t^2}((q_1-q_0)^2\\&+(q_2-q_1)^2+\cdots+(q_f-q_{N-1})^2)-\frac{i\gamma m^3}{\hbar\Delta t^3}((q_1-q_0)^4+(q_2-q_1)^4+\cdots+(q_f-q_{N-1})^4)\\&-\frac{if_0m}{\hbar\Delta t^2}\left[q_0(q_1-q_0)(\cos(\Omega\Delta t)-1)+\cdots+q_{N-1}(q_f-q_{N-1})(\cos(N\Omega\Delta t)-\cos((N-1)\Omega\Delta t))\right]\biggr]~.
\end{split}
\end{equation}
In the absence of the gravitational wave\cite{Pramanik}, the form of the propagator in eq.(\ref{1.44}) reads
\begin{equation}\label{1.45}
\left<q_f,T|q_0,0\right>=\sqrt{\frac{m}{2\pi i\hbar T}}e^{\frac{im}{2\hbar T}(q_f-q_0)^2}\left(1+\frac{3i\gamma m\hbar}{T}-6\gamma m^2\left(\frac{q_f-q_0}{T}\right)^2-\frac{i\gamma m^3}{\hbar T^3}(q_f-q_0)^4\right)~.
\end{equation} 
In presence of the gravitational wave, the propagator has the form given as
\begin{equation}\label{1.46}
\begin{split}
\left<q_f,T|q_0,0\right>&\simeq\sqrt{\frac{m}{2\pi i\hbar T}}e^{\frac{im}{2\hbar T}(q_f-q_0)^2}\biggr(1+\frac{3i\gamma m\hbar}{T}-6\gamma m^2\left(\frac{q_f-q_0}{T}\right)^2-\frac{i\gamma m^3}{\hbar T^3}(q_f-q_0)^4\\&+\frac{if_0mT}{\hbar}\left(\frac{(q_f-q_0)}{T}\right)^2[\cos(\Omega T)-1]-\frac{if_0q_f}{\hbar}\left(\frac{m(q_f-q_0)}{T}\right)[\cos(\Omega T)-1]\biggr)\\
&\simeq\sqrt{\frac{m}{2\pi i\hbar T}}\tilde{\mathcal{N}}(T,\gamma,f_0)e^{\frac{i}{\hbar}S_{cl}^{(f)}}
\end{split}
\end{equation}
where $S_{cl}^{(f)}$ is the classical action involving free particles and gravitational waves given by
\begin{equation}\label{1.47}
\begin{split}
S_{cl}^{(f)}=\frac{m}{2T}(q_f-q_0)^2-\frac{\gamma m^3}{T^3}(q_f-q_0)^4-\frac{mf_0}{2 T}(q_f-q_0)\left[(q_f\cos[\Omega T]-q_0)-(q_f-q_0)\frac{\sin[\Omega T]}{\Omega T}\right]
\end{split}
\end{equation}
and the form of the fluctuation term is given as follows
\begin{equation}\label{1.48}
\begin{split}
\tilde{\mathcal{N}}(T,\gamma,f_0)&\simeq 1+\frac{3i\gamma m\hbar}{T}-6\gamma m^2\left[\frac{q_f-q_0}{T}\right]^2+\frac{if_0mT}{\hbar}\left[\frac{q_f-q_0}{T}\right]^2[\cos(\Omega T)-1]-\frac{if_0mq_f}{\hbar}\left[\frac{q_f-q_0}{T}\right][\cos(\Omega T)-1]\\&-\frac{imf_0(q_f-q_0)}{2\hbar T}\biggr[\frac{(q_f-q_0)\sin(\Omega T)}{\Omega T}-(q_f\cos(\Omega T)-q_0)\biggr]~.
\end{split}
\end{equation}
\section{Summary}
In this work, we have constructed the path integral formalism of the propagation kernel for a resonant bar detector in the presence of a gravitational wave in the generalized uncertainty principle framework. In this framework, we have considered only quadratic order correction in the momentum. We have obtained the configuration space action for this system using the path integral formalism. With the action in hand, we have then obtained the equation of motion of the system. From the equation of motion, we observe that the overall frequency of the resonant detector shifts due to interaction with the gravitational wave. Next, we have used the form of the perturbation term for a circularly polarized gravitational wave to calculate the classical solution of the detector coordinate $q(t)$. Using this form of $q(t)$ we have finally obtained the classical action for a resonant bar detector interacting with a gravitational wave in the generalized uncertainty principle framework. We have then investigated the quantum fluctuation parameter of the bar detector in presence of a circularly polarized gravitational wave. In order to obtain the final form of the fluctuation, we have considered a free particle interacting with the gravitational wave. The final form of the fluctuation picks up correction terms due to both GUP correction and gravitational wave interaction. In this process we have neglected cross terms considering both GUP and GW interactions as it would result in a much smaller correction to the fluctuation factor than the other corrections present in the analytical form of the quantum fluctuation. It would also be important to carry out the above analysis in a linear GUP framework. However, we would like to report this in future. From the observational point of view, the importance of our work lies in the fact that resonant bar detectors have the potential of detecting gravitational waves with their present sensitivity at distances of the order of $10^2kpc$ from the earth. The propagator captures the quantum effects also. Hence, detectability of such quantum effects in resonant bar detectors is also a possibility in the near future. A knowhow of the propagator  of the detector coordinates is therefore necessary if not absolutely essential.

\end{document}